\documentclass[11pt,aps,prd,onecolumn,nofootinbib,superscriptaddress]{revtex4-2}
\usepackage[utf8]{inputenc}
\usepackage[T1]{fontenc}
\usepackage{amsmath,amssymb,amsfonts,bm,mathtools}
\usepackage{graphicx}
\usepackage{hyperref}
\usepackage{physics}
\usepackage{booktabs}
\usepackage{xcolor}
\hypersetup{colorlinks=true,citecolor=blue,linkcolor=blue,urlcolor=blue}
\usepackage{booktabs}   
\usepackage{tabularx}   
\usepackage{adjustbox}   
\usepackage{array}      
\usepackage{array}      
\usepackage{amsmath}    
\usepackage{multirow}

\begin{document}

\title{Beyond Plane Waves: Coherent Network Response to Collimated Gravitational-Wave Wavepackets}

\author{S. D. Campos}\email{sergiodc@ufscar.br}
\affiliation{Applied Mathematics Laboratory-DFQM/CCTS, Federal University of São Carlos, Rodovia João Leme dos Santos, km 110, CEP 18052780, São Paulo, Brazil}

\begin{abstract}
	We present a paraxial wavepacket model for structured, collimated gravitational-wave bursts and derive the coherent response of detector networks to these signals. For current terrestrial baselines such as LIGO–Virgo, analytic mismatch estimates confirm that the paraxial wavepacket model waveforms are effectively indistinguishable from standard sine–Gaussian bursts, validating the robustness of the plane-wave approximation in this regime. However, we identify a physical scaling regime relevant to third-generation networks and galactic-scale Pulsar Timing Arrays in which finite transverse structure—motivated by wave-optics lensing or ultra-relativistic beaming induces non-negligible geometric phase shifts. A toy event-level Monte Carlo study compares a standard burst-search ranking with a paraxial wavepacket model-constrained statistic that penalizes geometric inconsistencies across detectors. In this controlled setup, the model prior yields an illustrative factor of $\sim 3$–$4$ gain in detection efficiency at a fixed false-alarm rate, while maintaining performance on plane-wave-like signals. These results suggest that paraxial corrections may provide a necessary metrological framework for signal discrimination and unbiased parameter estimation in future cosmic-scale observations.
\end{abstract}

\maketitle

\section{Introduction}

The direct detection of gravitational waves by the LIGO and Virgo collaborations \cite{LIGO2016,Virgo2017} established interferometric gravitational wave (GW) astronomy as a precision probe of compact objects, strong-field gravity, and multimessenger astrophysics \cite{banerjee.2024,lu.2025,radice.2018,mcenery.2019,ando.2013, LIGO2016,Abbott2019_GWTC1,Abbott2017_GW170817,Abbott2017_MM,GW_MMA_Review}. Current analyses focus on a few signal classes: compact-binary coalescences, unmodeled bursts, stochastic backgrounds, and quasi-monochromatic continuous waves (see Ref. \cite{GW_MMA_Review} and references therein). In all cases, the detector response is typically modeled for an incoming plane wave, with the measured strain given by projecting the metric perturbation onto the interferometer tensor response \cite{cutler.2002}.

In the current ground-based interferometric GW framework (see Ref. \cite{cornish.2021} and references therein), kilometer-scale detectors, plane-wave signal models, and pipelines such as coherent WaveBurst \cite{klimenko.2016} and BayesWave \cite{cornish.2015} analyze transient signals using the local plane-wave approximation. The measured strain is obtained by projecting a spatially homogeneous metric perturbation onto the detector response tensor \cite{maggiore.book}. This approach has so far been robust and sufficient \cite{Abbott2019_GWTC1,abbott.2023,abbott.2019_allsky}: in the long-wavelength regime relevant to transient bursts, current methods describe all detected events efficiently and accurately, with any mismatch buried in detector noise.

As the field moves to next-generation observatories—space-based interferometers like LISA \cite{amaro.2017} and 3G ground networks such as the Einstein Telescope \cite{punturo.2010,abac.2025} and Cosmic Explorer \cite{hall.2022}—detections will cover much larger spatial and temporal scales. Over cosmic distances and baselines of thousands of kilometers, small deviations from an ideal plane wave may arise, even on meter scales. If a transient spin-2 disturbance is directionally collimated or has a finite transverse profile, it can induce subtle amplitude and phase corrections across an extended detector network.

This paper has three main goals. First, we show that, given current sensitivity limits and network configurations, standard burst templates (e.g., sine-Gaussians) are mathematically and practically optimal within the approximations and parameter ranges considered, fully capturing the physics accessible within the current detector noise. Second, we introduce a field-first phenomenological paraxial wavepacket model (PWM) for highly collimated structured GWs with a carrier frequency, finite duration, and explicit angular collimation in the paraxial regime. 

Finally, we show how this structured PWM is a theoretical extension for future cosmic-scale networks. Linking network observables to a unified beam-like geometry provides a criterion for when standard plane-wave descriptions suffice and when the finite-profile effects introduced here must be included. Thus, the PWM is a theoretically motivated refinement of the standard burst description, mainly relevant for future detector generations and quantitative studies of network-level distinguishability. 

Although the PWM developed here is intrinsically geometric and emission-mechanism independent, many astrophysical and cosmological scenarios naturally yield structured or strongly anisotropic GW profiles. Cosmic string cusps and kinks, for instance, produce highly collimated, beam-like GW bursts over cosmological distances \cite{damour.2000,key.2009,xia.2025}. In such cases, the idealized plane-wave approximation fails to capture the transverse spatial structure of the localized wavepacket.

Pronounced diffractive lensing, occurring when GWs pass near compact or small-scale mass inhomogeneities, can produce significant wave-optics corrections \cite{nakamura.1999,takahashi.2003}. These yield highly structured wavepackets that invalidate the homogeneous plane-wave approximation. By mathematically describing collimated wavepackets, the present framework provides a natural tool to quantify the spatial propagation and transverse dynamics of these transient structures.

The paper is organized as follows. Section \ref{sec:colimar} introduces the wavepacket description for a collimated GW pulse in linearized gravity. Section \ref{sec:response} details the physical scaling and detector response, extending the framework from single detectors and the LIGO/Virgo network to overlap-based distinguishability metrics and Pulsar Timing Arrays (PTAs). In Section \ref{sec:toy_mc_pwm}, we present an event-level toy Monte Carlo model to evaluate ranking statistics, showing how a PWM-constrained statistic and a geometric prior can reduce false positives or improve detection efficiency. Finally, Section \ref{sec:final} summarizes our main conclusions and outlines future directions, such as injection campaigns in realistic colored noise and integrating PWM-informed priors into coherent burst pipelines.

\section{Paraxial Wavepacket Formalism in Linearized Gravity}\label{sec:colimar}

\subsection{Covariant Formulation and Gaussian Envelopes}

In a realistic cosmological setting, the wavepacket propagates on a Friedmann–Robertson–Walker background, and its detector-frame parameters $(\omega_0,\sigma_t,h_0)$ relate to the source-frame values via standard cosmological redshift. Since the detector response depends only on the local metric perturbation near the instrument, our analysis consistently describes the wavepacket at the time of arrival, with all cosmological propagation effects absorbed into the effective parameter vector $\Theta$.


Consider the leading order of a small perturbation $h_{\mu\nu}$ of Minkowski spacetime, written as (hereafter $c=G=\hslash=1$)
\begin{equation}
	g_{\mu\nu}(x)=\eta_{\mu\nu}+h_{\mu\nu}(x),
	\qquad |h_{\mu\nu}|\ll 1,
\end{equation}
where Greek indices run from 0 to 3, and the background metric has signature $\eta_{\mu\nu}=\eta^{\mu\nu}=diagonal(-,+,+,+)$. In a vacuum and imposing the Lorenz gauge condition $\partial^\mu \bar h_{\mu\nu}=0$, the linearized Einstein equations can be written in terms of the trace-reversed perturbation as
\begin{equation}
	\bar h_{\mu\nu}=h_{\mu\nu}-\frac{1}{2}\eta_{\mu\nu}h,
	\qquad h=\eta^{\mu\nu}h_{\mu\nu},
\end{equation}
where
\begin{equation}
	\Box \bar h_{\mu\nu}=0,
\end{equation}
where $\Box=\eta^{\mu\nu}\partial_\mu\partial_\nu$ is the usual d'Alembertian. The physical propagating degrees of freedom are most transparently displayed in transverse-traceless (TT) gauge, where the perturbation named $h^{\mathrm{TT}}_{\mu\nu}$ has only two independent polarization states, conventionally denoted by $+$ and $\times$. In this gauge, looking for vacuum solutions, one has
\begin{eqnarray}
	h^{0\mu}=0,\, h_i^i=0,\, \partial_j h^{ij}=0,
\end{eqnarray}
where $i,j=1,2,3$. A plane-wave mode with a null wave-vector $k^\mu=(\omega,\mathbf{k})$, 
satisfying $k^\mu k_\mu=0$, takes the form
\begin{equation}
	h^{\mathrm{TT}}_{\mu\nu}(x)=\epsilon^A_{\mu\nu}(\hat{\mathbf{k}})
	e^{-ik\cdot x},
\end{equation}
where $A\in\{+ ,\times\}$ labels GW polarization. Moreover, the polarization tensor obeys the following conditions \cite{maggiore.book}
\begin{equation}
	\nonumber   k^\mu \epsilon^A_{\mu\nu}=0\,(\mathrm{transversality}),\qquad
	\eta^{\mu\nu}\epsilon^A_{\mu\nu}=0\,(\mathrm{traceless}),\qquad \mathrm{and}\qquad
	\epsilon^A_{\mu\nu}=\epsilon^A_{\nu\mu}\,(\mathrm{symmetry}),
\end{equation}
implying each component $h^{\mathrm{TT}}_{\mu\nu}$ satisfies a linear wave equation in flat spacetime. Therefore, we can choose a set of orthonormal tensors where
\begin{eqnarray}
	\epsilon_{\mu\nu}^A(\hat{\mathbf{k}})\epsilon^{B\,\mu\nu}(\hat{\mathbf{k}})=\delta^{AB}, \, A,B\in\{+,\times\}.
\end{eqnarray}

It is well-known that the plane wave solution for $h^{\mathrm{TT}}_A(x)$ has a general solution given by 
\begin{eqnarray}
	h^{\mathrm{TT}}_A(x)=\int \frac{d^4k}{(2\pi)^4}\tilde{h}_A(k)e^{-ik\cdot x},
\end{eqnarray}
and we also impose the massless condition: $k^2=0$. Integrating the above result, one obtains for the most general classical solution, expressed as a superposition of such modes, the following result
\begin{equation}
	h^{\mathrm{TT}}_{\mu\nu}(x)=
	\sum_{A=+ ,\times}
	\int \frac{d^3k}{(2\pi)^3 2\omega_{\mathbf{k}}}
	\left[a_A(\mathbf{k})\epsilon^A_{\mu\nu}(\hat{\mathbf{k}})e^{-ik\cdot x}
	+a_A^*(\mathbf{k})\epsilon^A_{\mu\nu}(\hat{\mathbf{k}})e^{ik\cdot x}\right],
	\label{eq:packet_general}
\end{equation}
where $\epsilon^A_{\mu\nu}$ represents the polarization states
of the wave (gravitons). Equation \eqref{eq:packet_general} gives the mode expansion of the most general TT solution of the vacuum linearized Einstein equations. The coefficients $a_A(\mathbf{k})$ encode the perturbation’s spectral distribution and angular dependence, while the polarization tensors $\epsilon_{\mu\nu}^A(\hat{\mathbf{k}})$ form a complete basis of transverse-traceless spin‑2 tensors for each propagation direction. Next, we choose a specific class of amplitudes $a_A(\mathbf{K})$ describing a collimated, narrowband wavepacket and derive its spacetime representation and detector response.


To represent a highly collimated structured GW pulse, one considers a wavepacket sharply concentrated around a preferred propagation direction $\hat{\mathbf{n}}$ and a preferred wavenumber $k_0$. Let $\mathbf{k}=k_\parallel \hat{\mathbf{n}}+\mathbf{k}_\perp$, with $\mathbf{k}_\perp\cdot\hat{\mathbf{n}}=0$. Concerning the goals of this work, a convenient Gaussian envelope can be written as
\begin{equation}
	a_A(\mathbf{k})=\mathcal{A}_A
	\exp\left[-\frac{(k_\parallel-k_0)^2}{2\sigma_\parallel^2}
	-\frac{|\mathbf{k}_\perp|^2}{2\sigma_\perp^2}\right]e^{i\varphi_A},
	\label{eq:momentum_env}
\end{equation}
which represents an envelope model incorporating the essential physical characteristics of a highly collimated pulse (see, for example, Ref. \cite{svelto.book}). The Gaussian dependence on $k_\parallel$ ensures a well-defined central angular frequency $\omega_0 = k_0$, 
together with a tunable longitudinal bandwidth $\Delta\omega \sim \sigma_\parallel$ 
which, via Fourier duality, corresponds to a finite temporal duration of the pulse. The field configuration has a Gaussian envelope modulating a monochromatic carrier wave. The global complex coefficients $\mathcal{A}_A e^{i\varphi_A}$ set the relative amplitudes and phases of the $+$ and $\times$ polarization components, allowing arbitrary linear or elliptical polarization states.

Similarly, the Gaussian dependence on $|\mathbf{k}_\perp|$ concentrates the support of $a_A(\mathbf{k})$ around the preferred propagation direction $\hat{\mathbf{n}}$, with an angular spread
\begin{eqnarray}
	\nonumber \Delta\theta \sim \frac{\sigma_\perp}{k_0}\ll 1,
\end{eqnarray}
that determines a finite transverse spatial extent $w \sim 1/\sigma_\perp$. In direct analogy with Gaussian laser beams, a collimated GW packet with transverse width $w$ and wavelength $\lambda_{\rm GW}$ may be associated with a Rayleigh length 
\begin{eqnarray}
	z_{\rm R} \sim \frac{\pi w^2}{\lambda_{\rm GW}}
\end{eqnarray} 
defined as the distance over which the beam radius spreads by a factor of $\sqrt{2}$ \cite{svelto.book}, beyond which the beam radius starts to grow appreciably. In the parameter range considered, $z_{\rm R}$ is much larger than any interferometer baseline, so the transverse width is effectively constant across the network. Using $f_0 \sim 10^2\,\mathrm{Hz}$, near the most sensitive band of current ground-based detectors\footnote{For definiteness, we will often quote numerical examples at a representative burst frequency \cite{abbott.2020}.} \cite{abbott.2020}, the GW wavelength is $\lambda_{\rm GW} \sim 3\times 10^6\,\mathrm{m}$. A beam with $w \sim 10^9\,\mathrm{m}$ then has a Rayleigh length $z_R \sim w^2/\lambda_{\rm GW} \sim 10^{11}\,\mathrm{m}$, about five orders of magnitude larger than typical interferometer baselines. In this regime, the GW wavepacket experiences negligible diffractive spreading through the network. Locally, the field is well approximated by a plane wave. Still, the global plane-wave approximation fails across the network because of the transverse-profile inhomogeneity on inter-site separations $B$, not longitudinal beam evolution.

Equations ~\eqref{eq:packet_general} and ~\eqref{eq:momentum_env} provide a fully covariant description of the proposed signal family. This formulation is intentionally independent of the production mechanism and can be viewed either as a classical field configuration or, in quantum field theory, as the classical limit of a coherent graviton state sharply peaked around a central mode \cite{maggiore.book}.

\subsection{Paraxial Regime and Wavefront Curvature}


For a narrowband wavepacket whose momentum-space support lies in a small cone about the propagation direction $\hat{\mathbf{n}}$, the significant propagation directions $\hat{\mathbf{k}}$ all satisfy $\angle(\hat{\mathbf{k}},\hat{\mathbf{n}})\ll 1$ \cite{svelto.book}. In this paraxial regime, the TT polarization tensors vary only weakly over the support of $a_A(\mathbf{k})$ and can be approximated by their value at the central direction, $\epsilon_{ij}^A(\hat{\mathbf{k}})\simeq e_{ij}^A(\hat{\mathbf{n}})$. This approximation is directly analogous to the standard treatment of polarization in paraxial optical beams \cite{maggiore.book,svelto.book}. Therefore, from the equation \eqref{eq:packet_general}, one obtains
\begin{eqnarray}
	h^{\mathrm{TT}}_{ij}(x)
	&=& \sum_{A}\int\!\frac{d^3k}{(2\pi)^3 2\omega_{\mathbf{k}}}\,
	a_A(\mathbf{k})\,\epsilon^A_{ij}(\hat{\mathbf{k}})\,e^{-ik\cdot x}
	+ \text{c.c.} \,\approx\,
	\sum_{A} e^A_{ij}(\hat{\mathbf{n}})\mathcal{H}_A(t,\mathbf{x})
\end{eqnarray}
where \text{c.c.} means the complex conjugate and with
\begin{equation}\label{eq:HA}
	\mathcal{H}_A(t,\mathbf{x})
	=
	\int\!\frac{d^3k}{(2\pi)^3 2\omega_{\mathbf{k}}}\,
	a_A(\mathbf{k})\,e^{-ik\cdot x}
	+ \text{c.c.}\,
\end{equation}

For a narrowband wavepacket, the dispersion relation can be linearized around the central mode
\begin{eqnarray}
	\omega(\mathbf{k}) \simeq \omega_0 + (\mathbf{k}-\mathbf{k}_0)\cdot\nabla_{\mathbf{k}}\omega|_{\mathbf{k}_0},
\end{eqnarray}
so that, along the propagation direction, one may write
\begin{eqnarray}
	\omega_{\mathbf{k}} \simeq \omega_0 + (k_\parallel - k_0)
\end{eqnarray}
and this implies that the phase $kx=\omega_\mathrm{k}t-\mathbf{k}\cdot\mathbf{x}$ depends fundamentally on the specific combination $u=t-\hat{\mathbf{n}}\cdot\mathbf{x}$. Now, neglecting the \text{c.c.} terms and separating the integration into transversal and longitudinal parts in equation \eqref{eq:HA}, one has
\begin{eqnarray}
	\nonumber \int dk_\parallel e^{-\frac{(k_\parallel-k_0)^2}{2\sigma_\parallel^2}} e^{-i(\omega_\mathrm{k}t-k_\parallel \hat{\mathbf{n}}\cdot\mathbf{x})}&\propto& e^{-\frac{u^2}{2\sigma_t^2}}e^{-i\omega_0 u},\\
	\nonumber \int d^2k_\perp e^{-\frac{|k_\perp|^2}{2\sigma_\perp^2}}e^{-\mathbf{k}_\perp\cdot\mathbf{x}_\perp}&\propto& e^{-\frac{|\mathbf{x}_\perp|^2}{2\omega^2}},
\end{eqnarray}
where $\sigma_t\sim 1/(c\sigma_\parallel)$ is the pulse duration, and $w\sim 1/\sigma_\perp$ is the transverse beam width. Then, the equation \eqref{eq:HA} takes the form of an envelope-modulated carrier
\begin{equation}
	\mathcal{H}_A(t,\mathbf{x})=
	h_{0,A}
	\exp\left[-\frac{u^2}{2\sigma_t^2}
	-\frac{|\mathbf{x}_\perp|^2}{2w^2}\right]
	\cos(\omega_0 u+\phi_A).
	\label{eq:spacetime_packet}
\end{equation}

Equation \eqref{eq:spacetime_packet} is formally equivalent to the paraxial Gaussian-beam solutions in classical wave optics, with a Gaussian envelope in retarded time and transverse spatial coordinates modulating a monochromatic carrier wave \cite{svelto.book}. The key difference lies not in the wavepacket structure but in the tensorial (spin-2) nature of the field and its coupling to matter through spacetime curvature, encoded in the geodesic deviation equation and detector response tensors. This contrasts with electromagnetism, where the coupling occurs via the Lorentz force on charged particles \cite{maggiore.book}. The resulting field takes the approximate form
\begin{equation}
	h^{\mathrm{TT}}_{ij}(t,\mathbf{x}) \approx
	\sum_{A=+ ,\times} e^A_{ij}(\hat{\mathbf{n}})
	\left[h_{0,A}
	\exp\left[-\frac{u^2}{2\sigma_t^2}
	-\frac{|\mathbf{x}_\perp|^2}{2w^2}\right]
	\cos(\omega_0 u+\phi_A) \right].
\end{equation}

At a single detector, in the strict plane-wave limit, evaluating equation \eqref{eq:spacetime_packet} at the detector position $\mathbf{x}_D$ gives
\begin{equation}
	\mathcal{H}_A(t,\mathbf{x}_D)
	=
	h_{0,A}
	\exp\left[-\frac{(t-t_0)^2}{2\sigma_t^2}\right]
	\cos\!\big[\omega_0 (t-t_0)+\phi_A\big],
\end{equation}
where $t_0 \equiv \hat{\mathbf{n}}\cdot\mathbf{x}_D$\,\,
and the transverse factors have been absorbed into $h_{0,A}$. For a single fixed polarization state and a single detector with an antenna pattern $F_A(\hat{\mathbf{n}},\psi)$ \cite{abbott.2009,abbott.2019_allsky}, the measured strain is
\begin{equation}\label{eq:strain1}
	h(t) = F_A(\hat{\mathbf{n}},\psi)\,\mathcal{H}_A(t,\mathbf{x}_D)
	= A\,\exp\left[-\frac{(t-t_0)^2}{2\sigma_t^2}\right]
	\cos\!\big[\omega_0 (t-t_0)+\varphi_0\big],
\end{equation}
which is precisely the standard sine-Gaussian pulse used in burst searches \cite{abbott.2009}.

The phenomenological packet in equation \eqref{eq:spacetime_packet} captures the key features of a collimated transient: finite duration and transverse extent. To quantify when the plane-wave approximation fails across a detector network, however, we must retain the full paraxial structure of the field, including the evolution of the wavefront curvature. Starting from the scalar envelope $H_A$ defined in equation \eqref{eq:HA}, one introduces the standard paraxial ansatz
\begin{equation}
	\mathcal{H}_A(\mathbf{x},t)
	= \Re \left\{\Psi_A(\mathbf{x})e^{i(k_0 z-\omega_0 t)}
	\right\},
\end{equation}
where $z=\hat n\cdot \mathbf{x}$ denotes the longitudinal propagation coordinate, and $\Psi_A$ is assumed to vary slowly compared with the carrier wavelength, 
\begin{equation}
	\left|
	\frac{\partial^2 \Psi_A}{\partial z^2}
	\right|
	\ll
	k_0
	\left|
	\frac{\partial \Psi_A}{\partial z}
	\right|.
\end{equation}

Substituting this ansatz into the vacuum wave equation,
\begin{equation}
	\Box \mathcal{H}_A = 0,
\end{equation}
and neglecting second longitudinal derivatives yields the paraxial equation
\begin{equation}
	2ik_0 \frac{\partial \Psi_A}{\partial z}
	+ \nabla_\perp^2 \Psi_A = 0,
\end{equation}
which is formally identical to the paraxial Helmholtz equation encountered in Gaussian-beam optics. Assuming $r = |\mathbf{x}_\perp|$, the fundamental axisymmetric solution is given by
\begin{equation}
	\Psi_A(r,z) =
	\Psi_{0,A} \frac{w_0}{w(z)}
	\exp \left[ -\frac{r^2}{w^2(z)}\right]
	\exp\left[i\frac{k_0 r^2}{2R(z)}\right]
	\left[-i\zeta(z) \right],
\end{equation}
where one uses the following definitions
\begin{eqnarray}
	w(z) = w_0 \sqrt{
		1+\left(\frac{z}{z_R}\right)^2 },\,\,
	R(z) = z\left[1+\left(\frac{z_R}{z}\right)^2\right],\,\,
	\zeta(z) = \tan^{-1}\left(\frac{z}{z_R}\right),
\end{eqnarray}
with Rayleigh length
\begin{equation}
	z_R = \frac{\pi w_0^2}{\lambda_{\rm GW}}
	= \frac{k_0 w_0^2}{2}.
\end{equation}

The complete transverse-traceless perturbation, therefore, becomes
\begin{equation}
	h_{ij}^{TT} = \sum_A e^A_{ij} \,\Psi_A(r,z)\,
	e^{i(k_0 z-\omega_0 t)}
	+ {\rm c.c.}
\end{equation}

The phase factor
\begin{equation}
	\Phi_{\rm curv}(r,z) = \frac{k_0 r^2}{2R(z)}
\end{equation}
represents the departure from a perfectly planar wavefront and encodes the beam's local curvature. For a detector network characterized by a baseline $B$, two sites probing different transverse positions within the packet experience a differential geometric phase
\begin{equation}
	\Delta\Phi \simeq \frac{k_0 B^2}{2R(z)}
	= \frac{\omega_0 B^2}{2cR(z)}.
	\label{eq:deltaphi}
\end{equation}

This quantity measures how well the plane-wave approximation holds across the network. For $\Delta\Phi\ll1$, the signal is effectively a conventional plane wave. As $\Delta\Phi$ approaches unity, wavefront curvature induces coherent phase differences between detectors, and the plane-wave description breaks down.


\section{Physical Scaling and Phase Corrections: From Terrestrial Networks to PTAs}\label{sec:response}

In the local plane-wave approximation, the measured strain in a terrestrial interferometer is obtained by projecting the (spatially homogeneous) TT metric perturbation onto the detector response tensor for a Michelson interferometer with orthogonal arm unit vectors $\hat{\mathbf{u}}$ and $\hat{\mathbf{v}}$ \cite{maggiore.book}. In the present case, the detector tensor is given by
\begin{equation}
	D^{ij}=\frac{1}{2}\left(u^i u^j-v^i v^j\right),
\end{equation}
so that the measured strain projected at the detector location $\mathbf{x}_I$ is written as
\begin{equation}
	h_I(t)=D_I^{ij} h^{\mathrm{TT}}_{ij}(t,\mathbf{x}_I),
	\label{eq:det_strain}
\end{equation}
where $I\in\{H,L,V\}$ denotes Hanford, Livingston, and Virgo. Substituting equation \eqref{eq:spacetime_packet} into the above projection gives
\begin{equation}
	h_I(t)=F_I^+(\hat{\mathbf{n}},\psi)\,\mathcal{H}_+(t,\mathbf{x}_I)
	+F_I^\times(\hat{\mathbf{n}},\psi)\,\mathcal{H}_\times(t,\mathbf{x}_I),
	\label{eq:pattern_response}
\end{equation}
where
\begin{equation}
	F_I^A = D_I^{ij} e^A_{ij}(\hat{\mathbf{n}},\psi)
	\qquad (A=+ ,\times)
\end{equation}
are the usual antenna pattern functions \cite{abbott.2009,abbott.2019_allsky}. The local-plane-wave approximation underlying equation \eqref{eq:pattern_response} is expected to be extremely accurate for current ground-based detectors, since the instrument size is tiny compared to the GW wavelength in the most sensitive band. Nevertheless, the formalism can still track finite-transverse-profile effects via the explicit dependence on $\mathbf{x}_I$.


If the beam width is much larger than the detector scale, the transverse factor may be evaluated at the detector center, so that the strain reduces to a delayed envelope-modulated carrier, written as
\begin{align}\label{eq:strain2}
	h_I(t) \simeq\,
	&F_I^+ h_{0,+} e^{-\frac{(t-t_I)^2}{2\sigma_t^2}}
	\cos\left[\omega_0(t-t_I)+\phi_+\right]\nonumber\\
	+&F_I^\times h_{0,\times} e^{-\frac{(t-t_I)^2}{2\sigma_t^2}}
	\cos\left[\omega_0(t-t_I)+\phi_\times\right],
\end{align}
with arrival time
\begin{equation}
	t_I=t_0+\hat{\mathbf{n}}\cdot \mathbf{x}_I.
\end{equation}

The Fourier transform of the result in \eqref{eq:strain2} is a narrowband wavepacket centered at $f_0 = \omega_0 / 2\pi$, with amplitude modulated by the detector antenna response. Without higher-order beam-pattern corrections, a single-detector wavepacket is indistinguishable from that of a suitably chosen elliptically polarized sine-Gaussian burst. In the long-wavelength limit, the strain in an individual detector is therefore indistinguishable from that of such a waveform, so any additional discriminatory power must come solely from network-level correlations in amplitude, phase, and polarization induced by the finite transverse spatial profile of the signal.

Although the plane-wave approximation is highly accurate for typical compact binary coalescences observed with current detector baselines, signals with strong directionality or significant waveform distortions may require more detailed modeling. For example, cosmic string networks produce highly collimated, beamed emission with energy concentrated along well-defined propagation directions; in this regime, the PWM yields a physically self-consistent and quantitatively accurate description of the spatial field profile \cite{seigman.book,bornwolf.book}. Likewise, when GWs undergo strong diffraction (wave-optics lensing) by dark matter clumps or compact objects, the resulting interference and magnification patterns generate structured wavefronts \cite{nakamura.1999,takahashi.2003}. In both scenarios, the transverse degrees of freedom in the PWM provide the necessary framework to characterize departures from spatial homogeneity.

\subsection{Coherent response of the LIGO/Virgo network}

In standard applications, the waveform parameters $\Theta$ are modeled as functions of the signal amplitude, central frequency, spectral bandwidth, propagation direction, and polarization, with additional phase or transverse spatial parameters introduced only when needed \cite{maggiore.book, LIGO2016, Abbott2019_GWTC1}. Here, for data analysis purposes, a practical parameter vector is defined as
\begin{equation}
	\Theta = \{h_0,\, \omega_0,\, \sigma_t,\, \hat{\mathbf{n}},\, \psi,\, \epsilon,\, w\},
\end{equation}
where $h_0$ denotes an overall amplitude scale, $\psi$ is the polarization angle, $\epsilon$ controls the relative mixing of the $+$ and $\times$ components, and $w$ may be replaced by the equivalent angular divergence $\Delta \theta$. In the limit $w\to\infty$, one recovers the usual plane-wave description with a Gaussian temporal envelope.

For a given parameter set $\Theta$, the measured data streams are given by
\begin{equation}
	s_I(t)=h_I(t;\Theta)+n_I(t),
\end{equation}
where $n_I(t)$ is detector noise. For a coherent transient observed in multiple interferometers, the key network observables are the relative arrival times $t_I - t_J$, the recovered amplitude and phase ratios after correcting for geometric delays, and the antenna response pattern $(F_H^A, F_L^A, F_V^A)$, which together determine the source sky location and polarization content \cite{maggiore.book}.

Because the Virgo detector is rotated relative to the nearly coaligned LIGO interferometers, its antenna response pattern complements those of Hanford and Livingston, improving sky localization and sensitivity to signal polarization \cite{schutz.2011}. For the signal family considered here, this configuration allows the network to test whether the transient is consistent with a standard plane-wave burst model or whether a more structured, beam-like parameterization better fits the data.


For known wavepacket parameters, the matched-filter inner product in detector $I$ is \cite{SathyaprakashSchutz,CreightonAnderson}
\begin{equation}\label{eq:innerprod}
	(a|b)_I = 4\Re\int_0^{\infty}\frac{\tilde a_I(f)\tilde b_I^*(f)}{S_{n,I}(f)}df,
\end{equation}
where $S_{n,I}(f)$ is the one-sided noise spectral density. The single-detector optimal signal-to-noise ratio (SNR) is
\begin{equation}
	\rho_I^2=(h_I|h_I)_I,
\end{equation}
and the coherent network SNR is approximately \cite{schutz.2011}
\begin{equation}
	\rho_{\mathrm{net}}^2 = \sum_I \rho_I^2,
\end{equation}
when cross-correlated noise can be neglected. This procedure systematically constructs a detectability map over the multidimensional parameter space $(h_0, f_0, \sigma_t, \hat{\mathbf{n}}, \psi, \epsilon, w)$. Detectability depends on the balance between antenna-pattern suppression and narrowband spectral concentration: a signal whose frequency content aligns with the detector’s most sensitive noise region can still be detected, even if geometric projection strongly reduces the observed strain in one or more interferometers.

\subsection{Distinguishability and data-analysis implications}\label{sec:distingue}


A natural first diagnostic is the overlap between the PWM and conventional burst templates. Given two candidate signals $h_1$ and $h_2$, the noise-weighted overlap in detector $I$ is \cite{brown.2012,schmidt.2024}
\begin{equation}\label{eq:innerprod2}
	\mathcal{O}_I(h_1,h_2)=
	\frac{(h_1|h_2)_I}{\sqrt{(h_1|h_1)_I (h_2|h_2)_I}},
\end{equation}
where the inner product is defined in equation \eqref{eq:innerprod}. One can define either detector-specific overlaps or a coherent network overlap by summing detector inner products. If the overlap between a highly collimated structured wavepacket and the optimal elliptically polarized sine-Gaussian template remains near unity over the relevant parameter space, the PWM is essentially a physically motivated reparameterization of an existing family of burst wavepackets. If, instead, the overlap systematically degrades in regions associated with finite-beam effects or correlated polarization structure, the PWM represents a fundamentally distinct target for search pipelines.

A coherent signal consistent with the PWM must reproduce: (i) the relative arrival times set by the propagation direction $\hat{\mathbf{n}}$; (ii) the detector-dependent amplitudes from the antenna pattern factors $F_I^A$; (iii) the phase relations from a common carrier frequency and pulse envelope; and, in principle, (iv) any residual dependence on the finite transverse profile through the detector positions $\mathbf{x}_I$. 

To show how the PWM serves as an intrinsic physical constraint against non-astrophysical transients, we compare its network-consistency criteria with those of traditional unmodeled standard burst searches (SBS). In a coherent network of $N_d$ detectors, the data vector $\mathbf{d}(t)$ is modeled as 
\begin{eqnarray}\mathbf{d}(t) = \mathbf{F}\mathbf{h}(t) + \mathbf{n},
\end{eqnarray}
where $\mathbf{F}$ represents the network antenna response matrix and $\mathbf{n}(t)$ is the instrumental noise \cite{harry.2008}. In standard unmodeled maximum-likelihood pipelines (such as coherent WaveBurst), the reconstruction uncouples the signal components at different sites \cite{klimenko.2005,klimenko.2008,klimenko.2016}, allowing the network to fit independent waveform degrees of freedom to maximize the standard coherent log-likelihood ratio $\ln \Lambda_{\text{SBS}}$. This unconstrained flexibility frequently allows localized, coincident instrumental glitches to be absorbed into the signal manifold \cite{blackburn.2008}. As mentioned previously, one proposes here that, conversely, the PWM restricts the search space to a rigid, physically motivated parameter vector $\Theta$, defining a penalized coherent likelihood function
\begin{equation}
	\ln \Lambda_{\text{PWM}} = \ln \Lambda_{\text{SBS}} - \frac{1}{2} \chi^2_{\text{geom}}(\Theta),
\end{equation}
where $\chi^2_{\text{geom}}(\Theta)$ quantifies the metric deviation of the reconstructed network data from the geometric constraints imposed by the paraxial transverse profile. This constraint directly impacts the network's null-stream vector, $\mathbf{x}_{\text{null}}(t) = \mathbf{M}_{\text{null}}\mathbf{d}(t)$, which projects the data onto the subspace orthogonal to the true signal manifold, such that $\mathbf{M}_{\text{null}}\mathbf{F} = \mathbf{0}$ \cite{gursel.1989,chatterji.2006,wen.2012}. For a physically realized paraxial burst characterized by the parameter vector $\Theta$, the residual energy present in the null stream is minimized, such that \cite{gursel.1989,chatterji.2006}
\begin{equation}
	E_{\text{null}} = \int \left| \mathbf{x}_{\text{null}}(t) \right|^2 dt \approx \int \left| \mathbf{M}_{\text{null}}\mathbf{n}(t) \right|^2 dt.
\end{equation}

This minimization underlies coherent consistency tests across multiple channels: the null stream cancels the GW strain tensor, thereby isolating residual non-Gaussian noise~\cite{wen.2012,sutton.2010}. In SBS pipelines, a terrestrial glitch can mimic an unmodeled burst because independent amplitude and phase adjustments at each site can artificially lower the null-stream energy. Under PWM, however, such uncoordinated instrumental variations cannot match the low-dimensional manifold set by $\Theta$, so a glitch incurs a large geometric penalty ($\chi^2_{\text{geom}} \gg 1$) and its energy leaks strongly into the null stream. 


To make the foregoing discussion more concrete, we consider the case of a single detector and a representative narrowband wavepacket characterized by the parameters
\begin{equation}
	\omega_0 = 2\pi \times 100~\mathrm{Hz},
	\qquad
	\sigma_t = 5~\mathrm{ms},
	\qquad
	h_{0,+} = h_{0,\times} = h_0,
	\label{eq:example_params}
\end{equation}
and take the plane-wave limit $w\rightarrow\infty$ so that transverse-profile effects can be neglected. In this regime, equation \eqref{eq:strain2} reduces to
\begin{equation}
	h_I^{\mathrm{pkt}}(t)
	\simeq
	A_I^{\mathrm{pkt}}
	e^{-(t-t_I)^2/2\sigma_t^2}
	\cos\!\big[\omega_0(t-t_I)+\phi_I^{\mathrm{pkt}}\big],
	\label{eq:example_packet_compact}
\end{equation}
where the antenna-pattern factors and polarization mixture determine the detector-dependent amplitude and phase. Of course, a reference sine-Gaussian in the same detector can be written as
\begin{equation}
	h_I^{\mathrm{SG}}(t)
	=
	A_I^{\mathrm{SG}}
	e^{-(t-t_I)^2/2\sigma_t^2}
	\cos\!\big[\omega_0(t-t_I)+\phi_I^{\mathrm{SG}}\big].
	\label{eq:example_sg_compact}
\end{equation}

If the detector noise varies slowly across the signal bandwidth, the noise-weighted inner product is well approximated by a time-domain Gaussian overlap. In the plane-wave limit, the packet waveform in equation \eqref{eq:example_packet_compact} should therefore have an overlap very close to unity with a suitably chosen reference sine-Gaussian. To see this, and dropping $I$ for clarity, one writes
\begin{eqnarray}
	h_1(t)&=&Ae^{-(t/\tau)^2}\cos (\omega_0 t),\\
	h_2(t)&=&Ae^{-(t/\tau)^2}\cos (\omega_0 t+\Delta\phi),
\end{eqnarray}
for the $A$, envelope width $\tau$, and carrier frequency $\omega_0$. Moreover, if the detector noise $S_n(t)$ is approximately constant across the narrow signal bandwidth, the noise‑weighted inner product reduces to a time‑domain Gaussian‑weighted inner product, and the normalized overlap equation \eqref{eq:innerprod2} becomes
\begin{eqnarray}\label{eq:innerprod3}
	\mathcal{O}=\frac{\int_{-\infty}^{\infty}h_1(t)h_2(t)dt}{\bigl[\int h_1^2(t)dt \int h_2^2(t)dt\bigr]^{1/2}}.
\end{eqnarray}

The second signal can be written as
\begin{eqnarray}
	\cos(\omega_0 t+\Delta\phi)=\cos(\omega_0 t)\cos(\Delta\phi)-\sin(\omega_0 t)\sin(\Delta\phi),
\end{eqnarray}
and the numerator in \eqref{eq:innerprod3} is given by
\begin{eqnarray}\label{eq:innerprod3}
	\int_{-\infty}^{\infty}h_1(t)h_2(t)dt=A^2\int e^{-2(t/\tau)^2}\cos(\omega_0 t)\bigl[\cos(\omega_0 t)\cos(\Delta\phi)-\sin(\omega_0 t)\sin(\Delta\phi) \bigr]dt,
\end{eqnarray}
where the integral over $\cos(\omega_0 t)\sin(\omega_0 t)$ vanishes by symmetry for a symmetric Gaussian centered at 0. Moreover, 
\begin{eqnarray}
	\left[\int h_1^2(t)dt \int h_2^2(t)dt\right]^{1/2}=\int h_1^2(t)dt=A^2\int e^{-2(t/\tau)^2}\cos^2(\omega_0 t)dt.
\end{eqnarray}

\begin{figure}[t!]
	\centering
	\includegraphics[width=0.7\columnwidth]{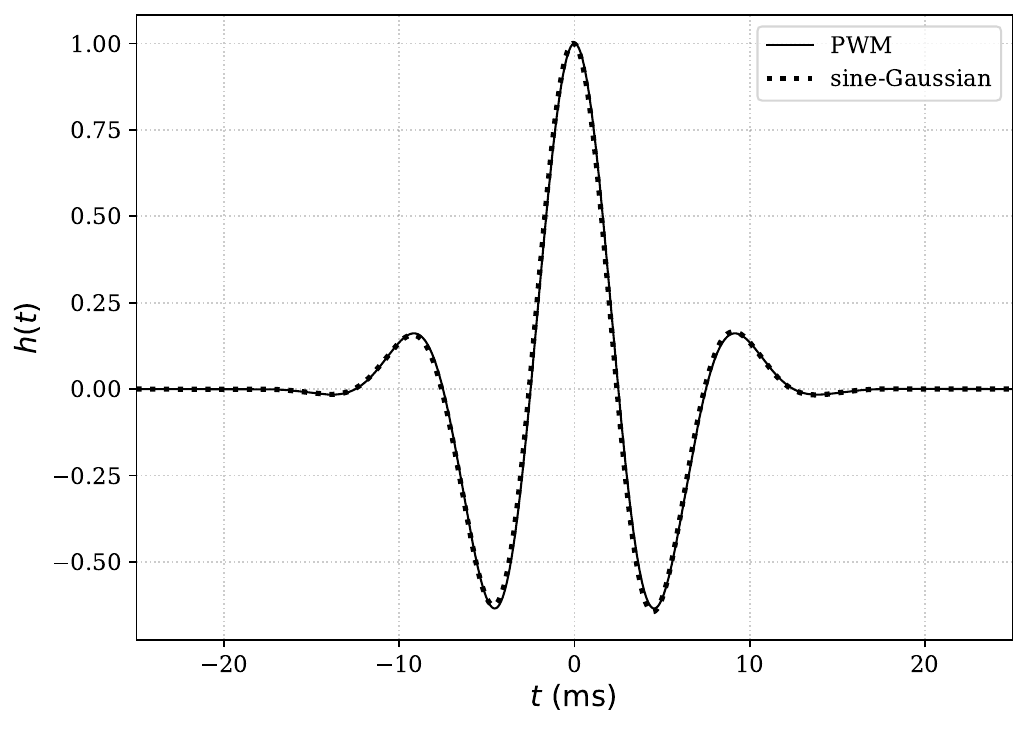}
	\caption{Strain in a single detector for a highly collimated GW (solid) and a reference sine-Gaussian burst (dotted), both with carrier frequency $f_0 = 100~\mathrm{Hz}$ and temporal width $\sigma_t = 5~\mathrm{ms}$. With a phase offset $\Delta\phi = 0.05$~rad, the single-detector overlap is $\mathcal{O}_I \simeq 0.9988$. The near coincidence of the curves shows that, in the plane-wave limit, the PWM is effectively indistinguishable from a suitably chosen elliptically polarized sine-Gaussian for a single detector.}
	\label{fig:packet_vs_sg}
\end{figure}

Therefore, all the amplitude and envelope integrals cancel, and we have the standard result for the inner product
\begin{eqnarray}\label{eq:signals}
	\mathcal{O}=\cos(\Delta\phi),
\end{eqnarray}
and a direct numerical evaluation using \eqref{eq:signals} confirms this expectation: for two signals with identical envelopes and a relative phase offset of $\Delta\phi = 0.05$~rad, one finds
\begin{equation}
	\mathcal{O}_I
	\equiv
	\mathcal{O}_I(h_I^{\mathrm{pkt}},h_I^{\mathrm{SG}})
	\simeq 0.9988,
	\label{eq:example_overlap_compact}
\end{equation}
while $\Delta\phi = 0.1$~rad gives $\mathcal{O}_I \simeq 0.995$. By contrast, pure amplitude differences at the 10\% level leave the overlap essentially unchanged at this level of approximation. Figure \ref{fig:packet_vs_sg} illustrates this point by comparing the time-domain strain of a highly collimated wavepacket with that of a reference sine-Gaussian sharing the same carrier frequency and temporal width but differing by a small phase offset.

\begin{figure}[t!]
	\centering
	\includegraphics[width=0.48\columnwidth]
	{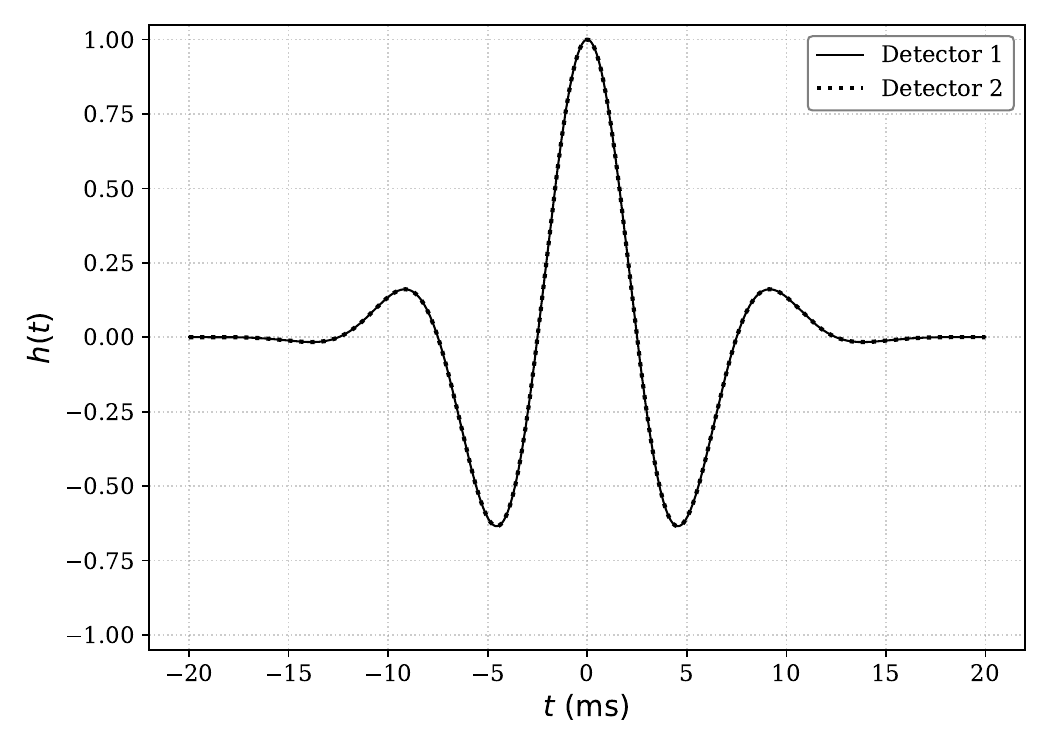}
	\includegraphics[width=0.48\columnwidth]
	{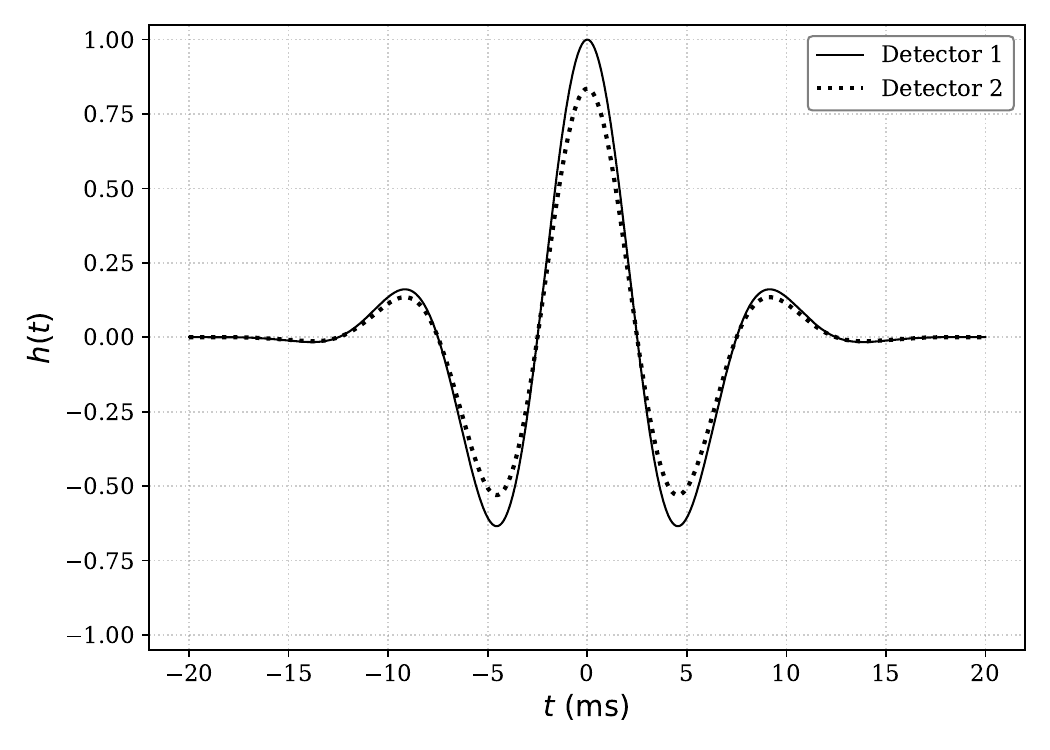}
	\caption{Strain in two interferometric detectors from the same collimated Gaussian GW packet. Left: realistic case with beam width $w = 10^9\,\mathrm{m}$ and transverse separation $|\Delta x_\perp| = 3\times 10^6\,\mathrm{m}$, giving $A_2/A_1 \simeq 0.999995$ so the traces are nearly identical. Right: illustrative case with $w = 5\times 10^6\,\mathrm{m}$ and the same separation, giving a clearly different response with $A_2/A_1 \simeq 0.84$.}
	\label{fig:prediction}
\end{figure}

However, a more pertinent question is whether a finite transverse spatial structure can induce small yet coherent deviations in the measured response across a network of detectors. To estimate that effect, let the beam width be finite but large, with
\begin{equation}
	w \sim 10^9~\mathrm{m},
	\label{eq:example_beamwidth}
\end{equation}
so that the beam remains much wider than an individual interferometer while no longer being parametrically infinite on the scale of intersite separations. In this case, the strain in detector $I$ retains a residual dependence on the detector position $\mathbf{x}_I$ through the transverse factor in equation \eqref{eq:spacetime_packet}. 

The scale $w$ does not necessarily refer to the entire beam size, but rather to the width of the interference fringes or near a caustic, where the amplitude changes rapidly over small distances \cite{takahashi.2003}. When Earth passes through one of these fringes, the transverse wave gradient becomes extremely large. For suitable source directions, the induced detector-to-detector phase shifts are naturally of order $10^{-2}$--$10^{-1}$~rad. Combined with the single-detector estimates above, this implies network-coherent mismatches at the $10^{-3}$--$10^{-2}$ level for otherwise identical sine-Gaussian templates. In other words,
\begin{equation}
	\mathcal{O}_{\mathrm{net}}
	\simeq
	1-\delta\mathcal{O}_{\mathrm{geom}},
	\qquad
	\delta\mathcal{O}_{\mathrm{geom}}
	\sim 10^{-3}-10^{-2},
	\label{eq:example_network_mismatch}
\end{equation}
for representative choices of $(\omega_0,\sigma_t)$ and sky location. Figure \ref{fig:prediction} shows the time-domain strain from the same collimated Gaussian GW packet at two interferometric detectors at different transverse positions in the beam. The small difference between the traces results from the finite transverse profile \(w\). Comparing the two panels shows that, for parameters relevant to current ground-based detectors, finite-transverse-profile effects are suppressed but still offer a controlled way to model geometric mismatches in more extreme setups. This illustrates network-level corrections that vanish in the plane-wave limit but can become relevant for large baselines.

To connect this geometric phase shift with the signal-analysis framework, consider the normalized overlap between the exact paraxial waveform and its plane-wave approximation. In this case, the inner product given by equation \eqref{eq:innerprod3} becomes
\begin{equation}
	O
	=
	\frac{(h_{\rm PWM}|h_{\rm PW})}
	{\sqrt{
			(h_{\rm PWM}|h_{\rm PWM})
			(h_{\rm PW}|h_{\rm PW})
	}}.
\end{equation}

Expanding the overlap for small phase differences,
$\Delta\Phi\ll1$, yields
\begin{equation}
	O
	\simeq
	1
	-
	\frac{\langle\Delta\Phi^2\rangle}{2},
\end{equation}
where the angular brackets denote the signal-weighted average over the waveform support. The corresponding mismatch is therefore
\begin{equation}
	M
	=
	1-O
	\simeq
	\frac{\langle\Delta\Phi^2\rangle}{2}.
	\label{eq:mismatch_paraxial}
\end{equation}

\begin{figure}[t!]
	\includegraphics[width=0.7\columnwidth]{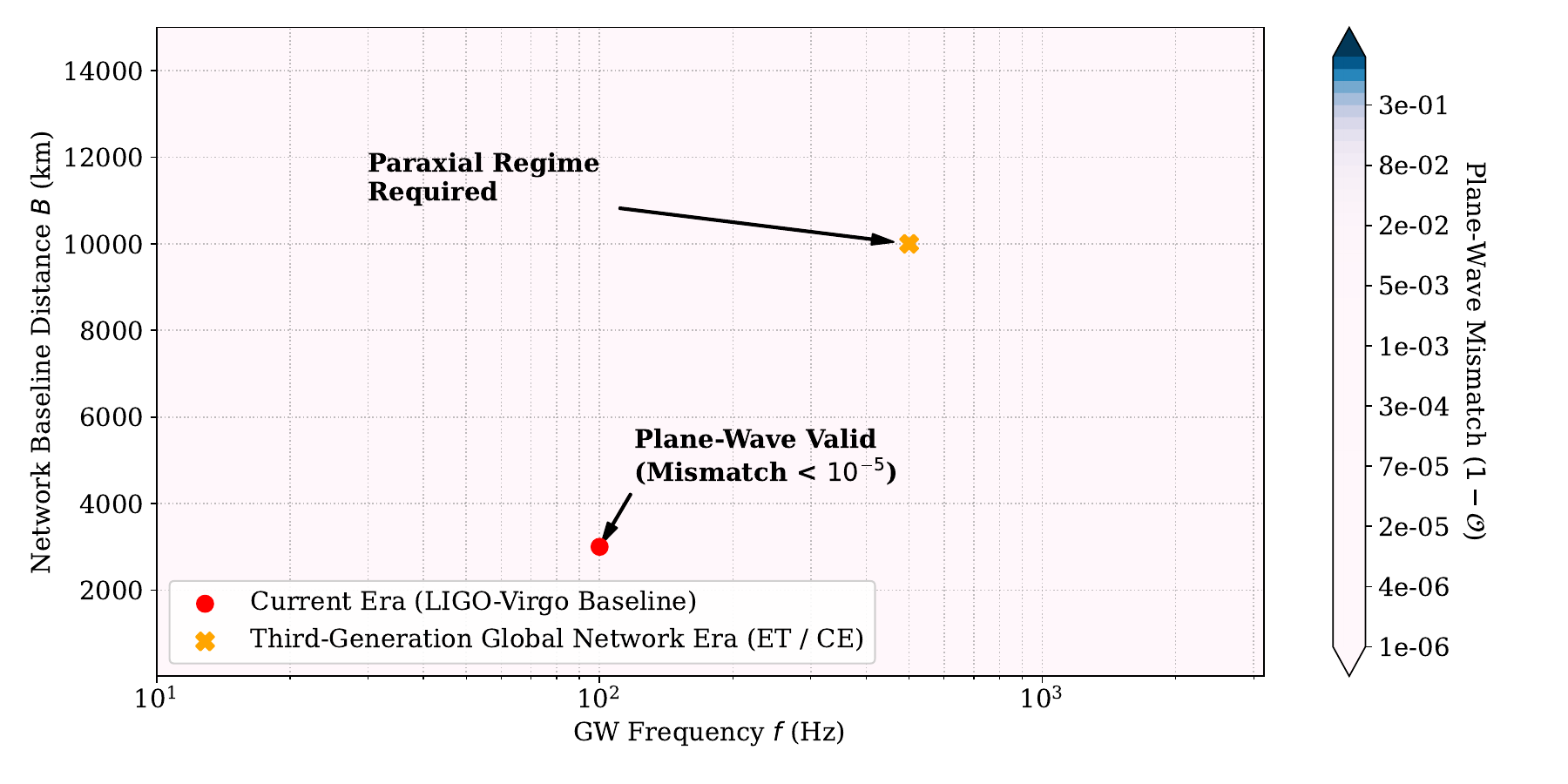}
	\caption{Wave–plane breakdown horizon versus network baseline $B$ and GW frequency $f$. Contours show the analytical mismatch ($1 - \mathcal{O}$) between a structured PWM and an SBS template. The red circle indicates the current terrestrial baseline regime (LIGO–Virgo), where SBS templates remain highly accurate. The orange cross indicates the future 3G regime, in which longer baselines and higher sensitivity enter the PWM, requiring structured templates to avoid systematic parameter biases.}
	\label{fig:wave_plane_horizon}
\end{figure}

Equation \eqref{eq:mismatch_paraxial} provides a physically motivated estimate for the onset of paraxial corrections and underlies the mismatch contours. Figure \ref{fig:wave_plane_horizon} shows the breakdown horizon across the ground-based detector band. In the current LIGO–Virgo era, the operational marker is a negligible mismatch ($< 10^{-5}$), confirming that the SBS (sine-Gaussian) model remains valid and effective. In contrast, 3G detectors, with longer effective baselines and higher-frequency sensitivity, enter the PWM regime. There, imposing a plane-wave template on a structured or beamed wavepacket yields mismatches that exceed typical precision-parameter-estimation thresholds, thereby establishing the proposed PWM as a metrological tool for 3G GW astronomy.

This quantitative threshold affects specific high-frequency astrophysical sources targeted by 3G observatories. As shown in Figure \ref{fig:wave_plane_horizon}, within $10^2\,\text{Hz} \le f \le 10^3\,\text{Hz}$ and for 3G-era global baselines ($B \sim 10^4\,\text{km}$), the plane-wave mismatch quickly exceeds $\mathcal{M} = 10^{-2}$ and grows to larger systematic deviations. This regime directly overlaps with expected emission from energetic, localized cosmological transients, such as high-frequency oscillations from cosmic string cusps \cite{damour.2000,abbott.2020} or structured wavepackets from sub-millisecond wave-optics lensing by compact dark matter clumps \cite{takahashi.2003,meena.2020}. Interestingly, cosmic strings emit gravitational waves at cusps moving at nearly $c$. The emission is directed into a very narrow cone with an opening angle of approximately $\theta \sim 1/(\gamma f)^{1/3}$ \cite{damour.2000}. When $\gamma$ is extremely high, the beam becomes so tightly focused that, despite traveling across cosmic distances, the beam's transverse size (its profile) remains small enough to be detected by a network of detectors spaced about $10^4$ km apart. For sources with ultra-relativistic Lorentz factors, intrinsic collimation limits beam expansion, so the paraxial approximation accurately describes the "edge" of the beam reaching Earth. Propagation through a field of random gravitational potentials (the "gas" of galaxies and dark matter) induces phase and amplitude corrections to the original wavefront, known as gravitational scintillation \cite{macquart.2004}. These fluctuations can create spatial (transverse) coherence scales on the order of $10^9$ to $10^{11}$ m.

\subsection{Scaling the Paraxial Response: From Terrestrial Networks to Pulsar Timing Arrays}\label{sec:pta}

One of the critical questions regarding the PWM is the magnitude of the geometric phase shift $\Delta\Phi$ compared to the standard plane-wave approximation. As derived in equation \eqref{eq:deltaphi}, the differential phase induced by the wavefront curvature between two detectors separated by a baseline $B$ is given by
\begin{equation}
	\Delta \Phi \simeq \frac{k_0 B^2}{2 R(z)} = \frac{\pi B^2}{\lambda_{GW} R(z)},
	\label{eq:phase_scaling}
\end{equation}
where $R(z)$ is the radius of curvature, which for astrophysical distances $D$ approaches the source distance ($R \approx D$).

For current and 3G ground-based interferometers, the baseline is limited by the Earth's geometry ($B \lesssim 10^4$ km). Considering a representative source at $D = 100$ Mpc and a carrier frequency of $f = 100$ Hz ($\lambda_{GW} \approx 3 \times 10^6$ m), the induced phase shift is given by
\begin{equation}\label{eq:pta1}
	\Delta \Phi_{\text{3G}} \sim \frac{\pi (10^7 \, \text{m})^2}{(3 \times 10^6 \, \text{m})(3 \times 10^{24} \, \text{m})} \sim 10^{-16} \, \text{rad},
\end{equation}
confirming that for terrestrial networks, the plane-wave approximation is exceptionally robust, as the curvature effects are suppressed by the small ratio of the baseline to the source distance.

However, the PWM framework becomes physically important for PTAs. In a PTA, the detector arms are defined by the distances between Earth and various pulsars in the galaxy, yielding galactic-scale baselines $B \sim 1$ kpc $\approx 3 \times 10^{19}$ m. PTAs are sensitive to nanohertz GWs ($f \sim 10^{-8}$ Hz), corresponding to wavelengths $\lambda_{GW} \sim 1$ pc $\approx 3 \times 10^{16}$ m. 

For a nearby supermassive black hole binary (SMBHB) at $D = 100$ Mpc, the paraxial phase correction scales as
\begin{equation}\label{eq:pta2}
	\Delta \Phi_{\text{PTA}} \sim \frac{\pi (3 \times 10^{19} \, \text{m})^2}{(3 \times 10^{16} \, \text{m})(3 \times 10^{24} \, \text{m})} \sim 10^{-2} - 10^{-1} \, \text{rad}.
\end{equation}

In this regime, the geometric phase shift is no longer negligible. A phase deviation of $O(10^{-2})$ radians is comparable to the timing residual precision required for GW detection in the nanohertz band. Furthermore, the transverse beam width $w$ in this context naturally scales to parsecs, characterizing structured emission from SMBHBs or diffraction patterns from galactic-scale lensing. Moreover, the 14 orders of magnitude difference between the results \eqref{eq:pta1} and \eqref{eq:pta2} justifies the use of PWM on galactic scales, even though it is optional on terrestrial scales.

As noted in previous studies on the breakdown of the plane-wave approximation for nearby PTA sources \cite{DengFinn2011}, including distance-dependent wavefront terms is essential for unbiased parameter estimation. The PWM developed here provides a self-consistent paraxial generalization of these effects, bridging the gap between local plane-wave templates and the full spherical-wave geometry required for galactic-scale GW astronomy.

\section{Event-Level Toy Monte Carlo and Ranking Statistics}
\label{sec:toy_mc_pwm}

To make the preceding discussion concrete, one can construct an event-level toy Monte Carlo (toy‑MC) to compare an SBS-like ranking statistic with one using a PWM geometric prior. This is not meant to replace injections into calibrated strain data or to reproduce the full flexibility of pipelines such as coherentWaveBurst or BayesWave. Its narrower goal is to test, in a controlled setting, whether restricting the signal manifold to a low-dimensional, physically motivated family can reduce the false alarm rate (FAR) or improve detection efficiency at fixed FAR.

The toy-MC is built from detector-level event summaries instead of time series. For each candidate event, it assigns reconstructed arrival times $t_I^{\rm rec}$, phases $\phi_I^{\rm rec}$, amplitudes $A_I^{\rm rec}$, and a network loudness proxy $\rho_{\rm net}$, where $I$ indexes the detectors. A simple SBS-like ranking statistic may then be defined here as
\begin{equation}\label{eq:sbs_stat}
	\mathcal{S}_{\rm SBS} = \rho_{\rm net} + \alpha \, C,
\end{equation}
where $C$ is a bounded coherence proxy built, for example, from the weighted phasor sum across detectors, and $\alpha$ is a tunable coefficient. In this construction, larger values of $\mathcal{S}_{\rm SBS}$ correspond to louder and more coherent events. The PWM-constrained statistic is defined by penalizing deviations from the best-fit geometric manifold associated with a single collimated wavepacket,
\begin{equation}
	\mathcal{S}_{\rm PWM} = \mathcal{S}_{\rm SBS} - \beta \, \chi^2_{\rm geom},
	\label{eq:toy_stat_pwm}
\end{equation}
where $\beta>0$ controls the strength of the penalty. A convenient event-level $\chi^2$, measuring how well a given candidate matches the PWM geometry, is written here as
\begin{eqnarray}
	\nonumber\chi^2_{\rm geom} = \sum_{I<J}
	\left( \frac{\Delta t_{IJ}^{\rm rec} - \Delta t_{IJ}^{\rm PWM}}{\sigma_t^{(IJ)}} \right)^2
	+ \sum_{I<J}
	\left( \frac{\Delta \phi_{IJ}^{\rm rec} - \Delta \phi_{IJ}^{\rm PWM}}{\sigma_\phi^{(IJ)}} \right)^2
	+ \\
	+\sum_{I<J}
	\left( \frac{\ln(A_I/A_J)^{\rm rec} - \ln(A_I/A_J)^{\rm PWM}}{\sigma_A^{(IJ)}} \right)^2,
	\label{eq:toy_chi2_geom}
\end{eqnarray}
where the time of arrival is given by $\Delta t_{IJ}=t_I-t_J$, the phases are defined by $\Delta\phi_{IJ}=\phi_I-\phi_J$, and the log-amplitude ratios are $\ln(A_I/A_J)^{\rm rec} - \ln(A_I/A_J)^{\rm PWM}$. At the same time, the PWM expectations are computed from the low-dimensional parameter vector $\Theta=\{h_0,\omega_0,\sigma_t,\hat n,\psi,\epsilon,w\}$ introduced previously. The resolution scales $(\sigma_t^{(IJ)}, \sigma_\phi^{(IJ)}, \sigma_A^{(IJ)})$ entering $\chi^2_{\rm geom}$ are treated here as hyperparameters chosen to place the toy-MC in a regime where geometric inconsistencies are of order unity for representative glitch populations; they should not be interpreted as directly inferred reconstruction uncertainties from any specific pipeline.

The background population is modeled as coincident, non-astrophysical transients. To test the PWM prior’s sensitivity to different glitch types, we define three background classes. The “incoherent” class uses events with independently drawn per-detector phases and amplitudes and only weakly correlated arrival times, representing highly unphysical coincidences. The “semi-coherent” class enforces a shared phase component and tighter temporal correlations, while still allowing detector-dependent phase and amplitude variations. The “coherent-like” class contains events whose temporal and phase evolution resemble a coherent signal, but whose amplitude ratios and phases violate a single-wavepacket geometric consistency relation across the network. Many such events can reach moderate network SNR and appear coherent in an SBS analysis, yet their geometry is generally incompatible with emission from a single collimated PWM.

The signal population has two classes. The first is plane‑wave–like coherent bursts with small inter‑detector phase and amplitude residuals, where standard sine‑Gaussian models are nearly optimal. The second is paraxial‑like coherent bursts with finite‑profile corrections and larger but strictly bounded deviations in inter‑detector phase, arrival time, and amplitude ratios. In both cases, reconstruction noise is added at the event-summary level, so comparisons occur at near-threshold rather than high–SNR.

\begin{figure}[t!]
	\centering
	\includegraphics[width=0.7\linewidth]{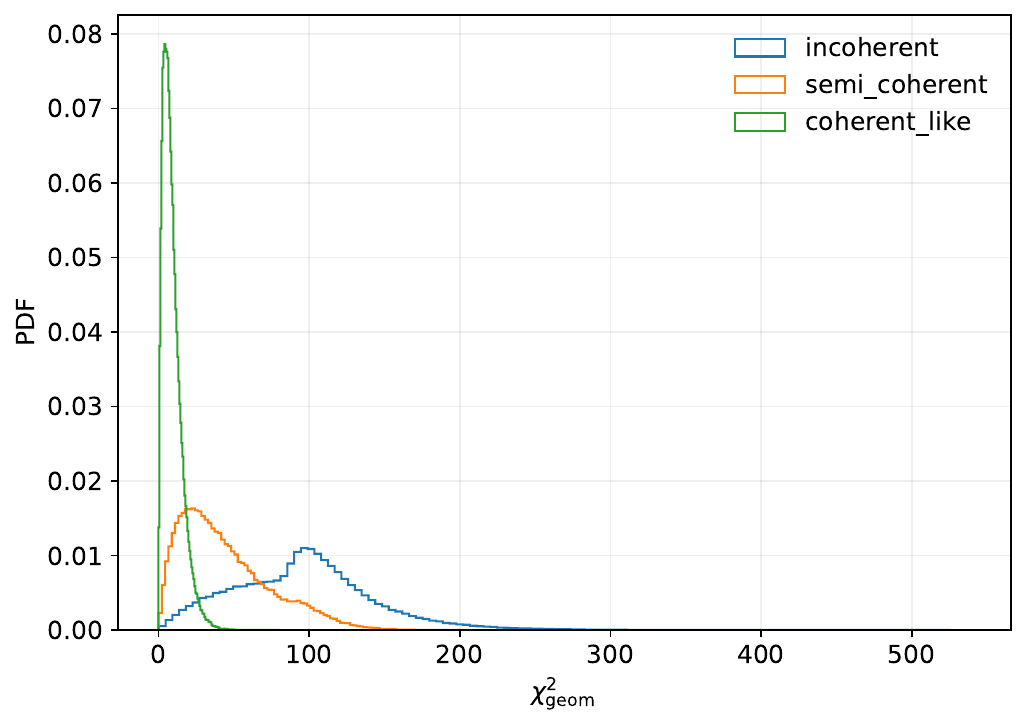}
	\caption{Distributions of $\chi^2_{\rm geom}$ for the three toy-MC background classes: incoherent, semi-coherent, and coherent-like glitches. Incoherent glitches occupy the highest $\chi^2_{\rm geom}$ values, semi-coherent glitches lie at intermediate values, and coherent-like glitches form a lower-$\chi^2$ tail that partially overlaps the signal region.}
	\label{fig:chi2_bg_classes}
\end{figure}

For each event in these populations, we compute the SBS-like and PWM statistics from equations \eqref{eq:sbs_stat} and \eqref{eq:toy_stat_pwm}, with $\Delta\chi^2 = \chi^2_{\rm plane} - \chi^2_{\rm PWM}$ defined via the geometric misfit in equation \eqref{eq:toy_chi2_geom}. Figure~\ref{fig:chi2_bg_classes} shows the resulting $\chi^2_{\rm geom}$ distributions for the three background classes: incoherent events occupy the largest $\chi^2_{\rm geom}$ values, semi-coherent events lie at intermediate values, and coherent-like glitches form a low-$\chi^2$ tail that partially overlaps the signal region. Injected plane-wave-like and paraxial-like signals drawn from the PWM manifold yield $\langle\chi^2_{\rm geom}\rangle \simeq 0.21$, while the combined background has $\langle\chi^2_{\rm geom}\rangle \simeq 96$ in this configuration. The additional mismatch $\Delta\chi^2$ is $\mathcal{O}(0.1)$ for plane-wave-like signals and $\mathcal{O}(1)$ for paraxial-like signals, reflecting that PWM matches the injections exactly. In contrast, the plane-wave hypothesis only approximates the paraxial class.

\begin{figure}[t!]
	\centering
	\includegraphics[width=\linewidth]{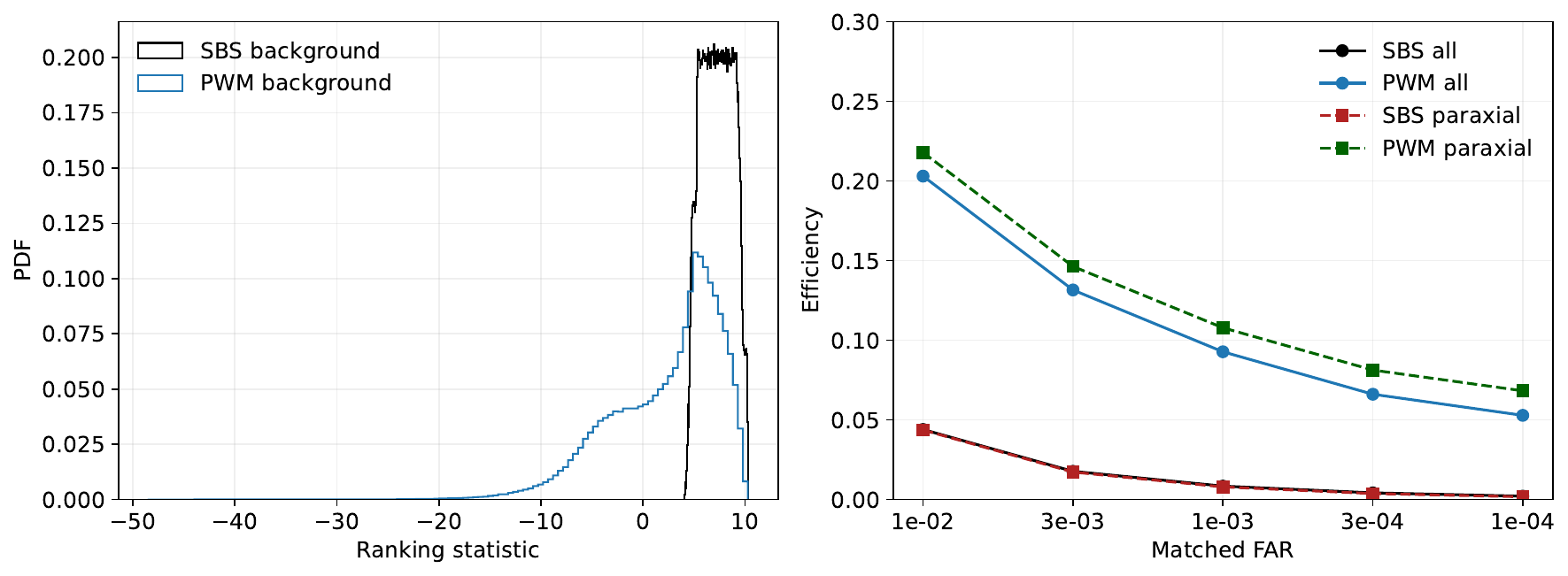}
	\caption{Toy-MC comparison between the SBS ranking statistic and the PWM-constrained statistic. Left: probability density functions of the background ranking statistics for SBS (black) and PWM (blue). Right: detection efficiency as a function of matched FAR for the same experiment. Solid curves show efficiencies for the full signal population, dashed curves for the paraxial-like subpopulation. For thresholds chosen to match the FAR of the corresponding SBS  background distribution, the PWM statistic recovers a factor of $\sim 3$--$4$ more signals at FAR levels between $10^{-2}$ and $10^{-4}$ in this controlled toy-MC, while the efficiencies for the paraxial-like subpopulation remain very similar for the two rankings.}
	\label{fig:toy_mc}
\end{figure}

The left panel of Figure \ref{fig:toy_mc} shows the background ranking-statistic PDFs for SBS (black) and PWM (blue). The strong geometric mismatch of the background under the PWM hypothesis shifts the $S_{\rm PWM}$ distribution to lower values than $S_{\rm SBS}$. In contrast, the signal distributions (not shown) remain sharply peaked at much higher ranks. In the right panel, thresholds are set to match the FAR of the SBS background. For these specific resolution scales $(\sigma_t,\sigma_\phi,\sigma_A)$ and background classes, the PWM statistic recovers $\sim 3$–$4$ times more signals than SBS at FARs between $10^{-2}$ and $10^{-4}$ in this toy Monte Carlo, for both the full signal set and the paraxial-like subset. It is important to stress that the numerical gain of $\sim 3$–$4$ in this configuration is therefore illustrative rather than universal, and depends on the adopted resolution scales, background mixture, and event-summarization scheme used in the toy-MC. 

\begin{figure}[t!]
	\centering
	\includegraphics[width=0.7\linewidth]{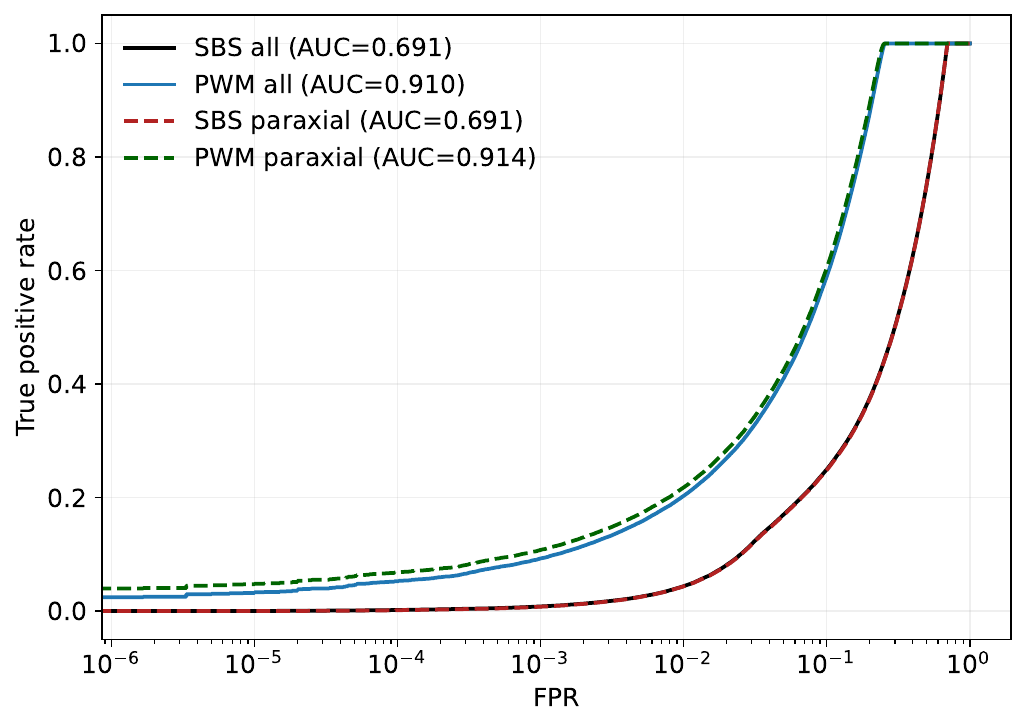}
	\caption{ROC curves for the SBS and PWM ranking statistics from the toy-MC study are shown for both the full combined signal population and the paraxial-like subsample. Over a broad range of FPR values, the PWM curves systematically lie above the SBS curves, indicating superior discrimination and ranking performance.}
	\label{fig:toy_mc_roc}
\end{figure}

The receiver operating characteristic (ROC) curves in Figure \ref{fig:toy_mc_roc} show the true positive rate versus the false positive rate (FPR)\footnote{Notice that $\mathrm{FAR}\simeq \mathrm{FPR}\times R_b$, where $R_b$ is the background trigger rate.} for the SBS and PWM test statistics. For both the full signal ensemble and the paraxial-like subset, the PWM-based curves consistently exceed those of SBS over a wide range of FPR, giving systematically larger areas under the curve (AUC). In this toy–MC study, adding the PWM-based geometric prior improves ranking performance: at any fixed FPR, a PWM-based search can, in principle, achieve higher detection efficiency than an SBS-based search on the same event summaries. We emphasize that the FPR shown on the ROC axis is defined as the fraction of background events that exceed a given threshold in this finite MC ensemble; it is related to, but distinct from, the astrophysical FAR, usually quoted as expected false triggers per unit observing time.

These gains arise in a deliberately idealized setting and depend on the chosen resolution scales $(\sigma_t,\sigma_\phi,\sigma_A)$, the background model, and the glitch-class mix. Our goal is not to reproduce the full behavior of pipelines like coherentWaveBurst or BayesWave, but to isolate and illustrate how a low-dimensional, physically constrained PWM manifold can enhance coherent ranking amid glitch-like coincidences, even when using only detector-summary variables. A natural next step is to impose PWM constraints in realistic injection campaigns and full end-to-end analyses with existing burst pipelines to measure how much of the toy-MC gain survives under more realistic noise and glitch populations.

\section{Discussion and Final Remarks}\label{sec:final}

The PWM developed in this work provides a theoretically motivated refinement of the standard plane-wave burst description, confirming that for current terrestrial sensitivities and baselines, conventional sine–Gaussian templates remain mathematically optimal, as departures from an ideal plane wave remain far below the noise floor \cite{abbott.2023,abbott.2019_allsky,abbott.2020,abbott.2009}. In this regime, the PWM is effectively indistinguishable from elliptically polarized sine–Gaussian bursts, as shown by the high overlaps in Sec.~\ref{sec:response} and the nearly identical performance of PWM and SBS rankings for plane-wave-like signals in our toy Monte Carlo study.

However, the PWM's main utility lies in its ability to impose physical constraints on coherent amplitudes, phases, and polarizations across detector networks. By encoding signal properties into a single low-dimensional parameter vector $\Theta$, analogous to Gaussian-beam optics \cite{maggiore.book,svelto.book}, this framework reduces the effective dimensionality of the signal manifold explored by pipelines such as coherent WaveBurst and BayesWave \cite{cornish.2021,klimenko.2016}. The toy-MC in Sec.~\ref{sec:toy_mc_pwm} shows explicitly that such geometric priors may improve detection efficiency by a factor of $\sim 3$--$4$ at fixed FAR by creating a strong separation in geometric misfit $\chi^2_{\rm geom}$ between signals and uncoordinated instrumental glitches.

Looking forward, the PWM extends GW modeling into regimes where the plane-wave approximation is expected to break down. As inter-detector baselines grow and sensitivities improve in 3G observatories and space-based missions like LISA \cite{amaro.2017,punturo.2010,abac.2025,hall.2022}, networks become sensitive to subtle geometric phase shifts. Our analysis indicates that while current LIGO-Virgo mismatches are negligible ($1 - O \lesssim 10^{-5}$), 3G configurations may encounter systematic mismatches of $10^{-3}$--$10^{-2}$. More critically, our scaling analysis in Sec.~\ref{sec:response} shows that for galactic-scale observations such as PTAs, paraxial corrections are essential for accurate parameter estimation due to the interplay between kiloparsec baselines and nanohertz wavelengths.

Beyond its statistical advantages, the PWM offers a geometric framework for modeling anisotropic or beamed sources. Cosmic string cusps \cite{damour.2000,key.2009,xia.2025} and wave-optics lensing by compact dark matter inhomogeneities \cite{nakamura.1999,takahashi.2003} naturally yield structured wavepackets that violate spatial homogeneity. By parameterizing these transients with the compact vector $\Theta$, this model provides a direct link between abstract signal manifolds and physically interpretable source classes. This is essential for the multi-messenger era, where joint GW, electromagnetic, and neutrino observations require unified models that encode both geometric propagation and source physics \cite{banerjee.2024,lu.2025,radice.2018,mcenery.2019,ando.2013}.

In conclusion, while this work does not claim immediate sensitivity gains for current networks, it establishes the statistical and geometric basis needed to determine when structured wavepacket models should be included in the GW astronomy toolkit. Future work should focus on source-specific PWM implementations and the integration of paraxial priors into existing coherent pipelines to measure how much of the toy-MC efficiency gains survive in the presence of non-Gaussian noise and realistic glitch populations.

\section*{Acknowledgments}

SDC acknowledges the Federal University of São Carlos (UFSCar) and the Applied Mathematics Laboratory (DFQM/CCTS) for their institutional support.

\end{document}